\documentclass[epj]{svjour}

\usepackage{latexsym}
\usepackage{graphicx}
\usepackage{fancyhdr}
\def\makeheadbox{{
 }}
\def\mytitle{My title} 
\def\myauthors{My name}  
\def\mytype{My type of session}
\def\mysession{My session}

\def\mytitle{Charged Higgs production in the flavored MSSM} 
\def\myauthors{Michael Spannowsky}    
\def\mytype{Contributed Talk}
\def\mysession{Colliders - Higgs Phenomenology}

\begin{document}
\title{Charged Higgs production in the flavored MSSM}
\author{Michael Spannowsky\inst{1}
\thanks{\emph{Email:} msp@particle.uni-karlsruhe.de}%
}                     
%
%
\institute{Institut f\"ur Theoretische Physik, Universit\"at Karlsruhe, D-76131 Karlsruhe, Germany}
%
%
\abstract{In the Minimal Supersymmetric Standard Model squark mixing can significantly change the pattern of charged-Higgs production. We find that non-minimal flavor structures can have a sizeable impact on the charged-Higgs production cross section, whereas charged-Higgs searches may be able to probe flavor structures not accessible to rare kaon, bottom or charm experiments, and can invalidate the assumption of minimal flavor violation.
} 
\maketitle
%

\section{Introduction}
\label{intro}

The main task in high-energy physics over the coming years is to understand the nature of electroweak symmetry breaking. In the Minimal Supersymmetric Standard Model (MSSM) two Higgs doublets are needed to give mass to all fermions. Such an extended model with each Higgs doublet coupling exclusively to up-type or down-type fermions is generally referred to as a two-Higgs-doublet model (2HDM) of type~II. It predicts two $CP$-even neutral Higgs bosons ($h^0$ and $H^0$), one $CP$-odd neutral Higgs boson ($A^0$) and a charged Higgs boson pair ($H^{\pm}$). 
Because of the absence of a $H^{\pm}W^{\mp}Z$ vertex at tree level, the production of a single charged Higgs and its decay operate foremost via Yukawa couplings, whereas the heavy-quark Yukawa couplings to the heavy Higgs states are governed by $y_b \tan \beta$ and by $y_t/\tan\beta$. Unfortunately, for $2 \leq \tan\beta \leq 20$ the rates are small and no promising detection channel is known \cite{CMScol}.

Within the Standard Model flavor symmetry breaking is governed solely by the Yukawa interactions, the spurions of flavor symmetry breaking. Applying this concept to extensions of the Standard Model leads to the notion of minimal flavor violation (MFV)~\cite{mfv}:
in an MFV model there are no other sources of flavor
violation than the Yukawa couplings. However, general soft SUSY breaking introduces new sources of flavor violation. In MFV (i) all soft scalar squark masses need to be diagonal in flavor space and (ii) all tri-scalar $A$-terms describing the squark--squark--Higgs couplings have to be proportional to the Yukawas.

\begin{eqnarray}
\label{eq:soft} 
\mathcal{L}_{\rm soft} 
  = &-\tilde U^* m_{\tilde U}^2 \tilde U - \tilde D^* m_{\tilde D}^2 \tilde D - \tilde Q^\dagger m_{\tilde{Q}}^2 \tilde Q \nonumber\\
   &-  \left[ \tilde Q \bar A^u \tilde U^* H_u - \tilde Q \bar A^d \tilde D^* H_d  + \mathrm{h.c.} 
      \right].  
\end{eqnarray}

The hermitian $6 \times 6$ squark mass matrices ${\mathcal M}_q^2$ for 
up and down-type squarks collect $D$, $F$ and soft terms from the SUSY breaking Lagrangian Eq.(\ref{eq:soft}). They are composed out of the left and right-handed blocks $M^2_{q \, AB}$. Each block is a $3 \times 3$ matrix in generation space:
\begin{equation}
{\mathcal M}_q^2 =\left( \begin{array}{cc}
M^2_{q \, LL} & M^2_{q \, LR} \\ 
M_{q \, LR}^{2 \, \dagger} & M^2_{q \,RR}
\end{array} \right)(q=u,d; A,B=L,R) \; .
\label{eq:mass_matrix}
\end{equation}
The explicit expressions for the $M^2_{q \, AB}$ are given in \cite{ourpaper}.
To trace back and discuss the sources of new--physics flavor violation, 
it is useful to define the dimensionless mass insertions 
 $\delta_{AB,ij}^q \equiv \frac{M^2_{q \, AB \, ij}}
                              {\tilde m^2}$ \cite{Hall:1985dx,fcnc-susy}.
The denominator is the geometric mean $\tilde{m}^2 = m_{A \, ii} m_{B \,jj}$
of the squared scalar masses of 
$\tilde q_{A i}$ and $\tilde q_{B j}$.
The off-diagonal entries of
$\delta_{AB}^q$ are significant only in non-MFV models 
and
can be complex, inducing $CP$ violation. We 
confine ourselves to real $\delta_{AB}^q$.
Note that in our numerical calculations we 
diagonalize the squark mass matrices and do not
employ a perturbative expansion in the $\delta_{AB}^q$, avoiding the 
calculation of the squark unitary transformations~\cite{Hall:1985dx}.

\section{Constraints on parameter space}\label{sec:constraints}

Especially flavor physics can strongly constrains the free parameters from the soft-breaking lagrangian, relevant for the enhancement of the charged-Higgs production. Flavor violation among down squarks is more severely constrained, because in $K$ and $B$ physics down-squark effects can be mediated by strongly interacting gluino loops, while up-squark effects are mediated by the weak interaction. Furthermore, mixing between first and second ge\-ne\-ra\-tion squarks is tightly con\-strain\-ed by $K$-physics~\cite{fcnc-susy,Colangelo:1998pm} and by the recent measurements of $D^0 \bar D^0$-mixing~\cite{Nir:2007ac}. Hence, we can limit our analysis to up-squark mixing between the first/second and third generation while neglecting down-squark mixing, as long as it is not required by $SU(2)$. Particularly constraining are the radiative decays $B \to X_s \gamma$ \cite{b_s_gamma_ex,b_s_gamma_th} and $B \to \rho \gamma$ \cite{rhogamma}, 
the semileptonic decays $B \to X_s \ell^+ \ell^-$ \cite{b_s_gamma_th,bsll,susy-Zpenguin} and $B \to \pi \ell^+ \ell^-$ \cite{B2pill}, 
and the $B_{d,s}$ ${-}$ $\overline{B}_{d,s}$ mass differences \cite{bs_mix,Bertolini:1990if,mixing-th}. All constraints on the supersymmetric flavor sector we implement at 90~\% C.L.:
\begin{itemize}
\item $A_{ii}^{u,d}$: diagonal $A$-term entries contribute
  to up and down-quark masses at one loop. We require perturbativity of SUSY-QCD corrections 
  $\delta m_q \leq m_q$.   
\item $A_{33}^{u,d}$: loop corrections 
  lift the lighter MSSM Higgs mass from $m_Z$ 
  to above the LEP2 limits.
\item $A_{13}^{d},A_{23}^{d}$, $A_{31}^{d},A_{32}^{d}$: general vacuum
  stability constraints limit the inter-generational $A$-terms~\cite{vacuum}. 
\item $A_{23}^{u,d}, (m_{\tilde U_{L}, \tilde D_{L,R}}^2)_{23}$:
 constrained by ($b \to s$)-type measurements. 
  $\overline{B}_s{-}B_s$ mixing mass difference $\Delta m_s$ implies $\frac{\Delta m_s}{\Delta m_s^{\rm SM}} = 1.00 \pm 0.44.$
  For the semileptonic and radiative decays we demand $2.94 \cdot 10^{-4} < {\rm BR} (B $ $\to$ $ X_s \gamma) < 4.14 \cdot 10^{-4}$ and $2.8 \cdot 10^{-6}< {\rm BR}(B \to X_s \ell^+ \ell^-) < 6.2 \cdot 10^{-6}$.
\item $A_{13}^{u,d}, (m_{\tilde U_{L}, \tilde D_{L,R}}^2)_{13}$:
  similarly, mixing between the first and third generation in the up and
  the down sector is constrained by $b \to d$ transitions: $\frac{\Delta m_d}{\Delta m_d^{\rm SM}} = 1.00 \pm  0.54$, $0.63 \cdot 10^{-6} <{\rm BR}(B \to \rho \gamma) < 1.24 \cdot 10^{-6}$ and ${\rm BR}(B \to \pi \ell^+ \ell^-) <9.1 \cdot 10^{-8}$.
\item $m_{\tilde{U}_{L}}^{2}$ and $m_{\tilde{D}_{L}}^{2}$: because
  SUSY breaking respects the $SU(2)$ gauge symmetry, the doublet
  soft-brea\-king masses are identical. In the super-CKM basis
  this means $m_{\tilde{U}_{L}}^{2}=V\cdot m_{\tilde{D}_{L}}^{2}\cdot V^{\dagger }$.
\item 
  Inter-generational mixing involving the third generation always
  affects the lightest Higgs mass and the $\rho$
  parameter~\cite{top_fcnc}.
\item Tevatron searches for mass-degenerate first- and second-generation squarks require $m_{\tilde q} > 200$ GeV \cite{tev_limits}.
\end{itemize}
The corresponding mass-matrix entries
$A^u_{3i}$ and $m^2_{\tilde U_{R} \, i3}$ are only very loosely bounded by 
flavor physics\footnote{We strictly use the convention
$A_{ij} = A_{L_i R_j} \neq A_{ji}$.}.
The reason is that they involve right-handed squarks $\tilde u_R$
and $\tilde c_R$; those enter FCNC processes with external 
down quarks only via higgsino vertices proportional to the small up and charm 
Yukawa. Hence, the  $\delta^u_{LR \, 3 i}$ and $\delta^u_{RR \, i3}$, ($i=1,2$)
are currently the least constrained flavored SUSY couplings. In the following we investigate 
the potential impact of these relevant $\delta^u_{3 i}$ on charged-Higgs collider searches.

\section{Single-Charged-Higgs Production}
\label{sec:singleH}

We start by considering single-charged-Higgs production at tree level from quark--antiquark
scattering, $q\bar q^\prime$ $\to H^\pm$, at the LHC. The amplitude for single-Higgs production
in the type-II two-Higgs-doublet model is proportional to the quark
Yukawa, thus small unless third-generation quarks are
involved \cite{single_higgs}.  This chiral suppression is generic and with proper
assumptions survives radiative corrections. For $\tan\beta = 7$ and a charged-Higgs mass of $m_{H^\pm}= 188~ \mathrm{GeV}$ the $H^+$ production cross section at the LHC in the 2HDM is $41.2~\mathrm{fb}$, using the $\overline{\rm MS}$ quark masses given in \cite{ourpaper}. 

The irreducible background to
our searches is single-$W$ production, $q\bar q^\prime \to W^\pm$. The
$W^+$ production cross section of $90 \cdot 10^6~\mathrm{fb}$
will be a serious challenge to our $H^+$ search in the
two-Higgs-doublet model.

Not assuming MFV has serious impact on the production rate
for $q \bar q^\prime \to H^\pm$. Squark loops will weaken the CKM
suppression at the charged-Higgs--bottom vertex through flavor mixing.
The dominant one-loop 
corrections are due to gluino vertex and self-energy diagrams.
Beyond MFV, 
the loop diagrams do not have to include a (Dirac)
quark mass to ensure the chiral limit of the theory.
Instead, we can for example combine a Majorana mass with a 
left-right mixing $\delta_{LR}$ among the squarks. This combination can lift the supersymmetric
charged Higgs
production rate above the two-Higgs-doublet model 
prediction, despite its loop suppression.
The MSSM Lagrang\-ian we define at the
weak scale, so all parameters are evaluated 
at the scale of the charged Higgs mass.

\begin{figure}[t]
 \begin{center}
   \includegraphics[width=0.24\textwidth,height=0.22\textwidth,angle=0]{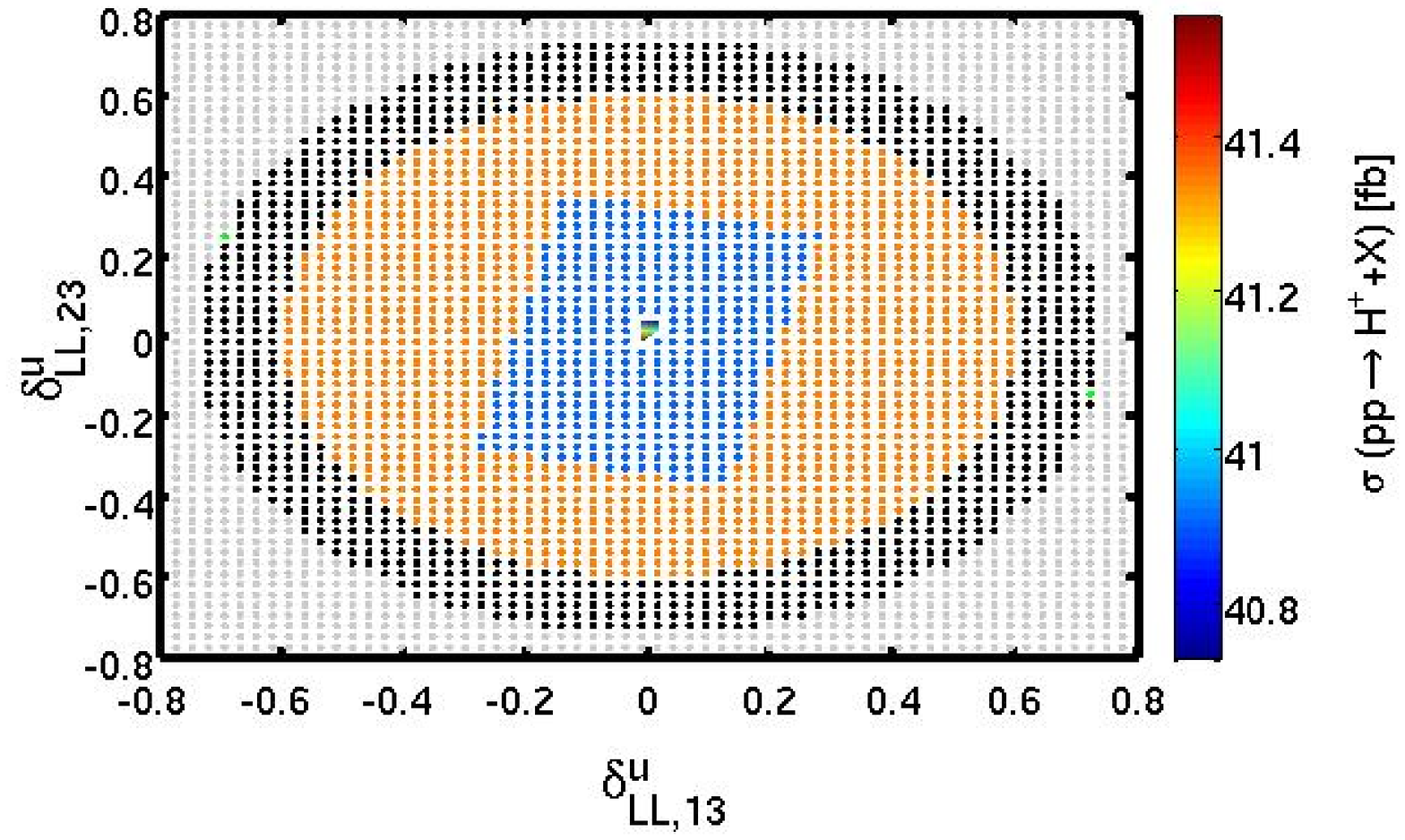} 
   \includegraphics[width=0.24\textwidth,height=0.22\textwidth,angle=0]{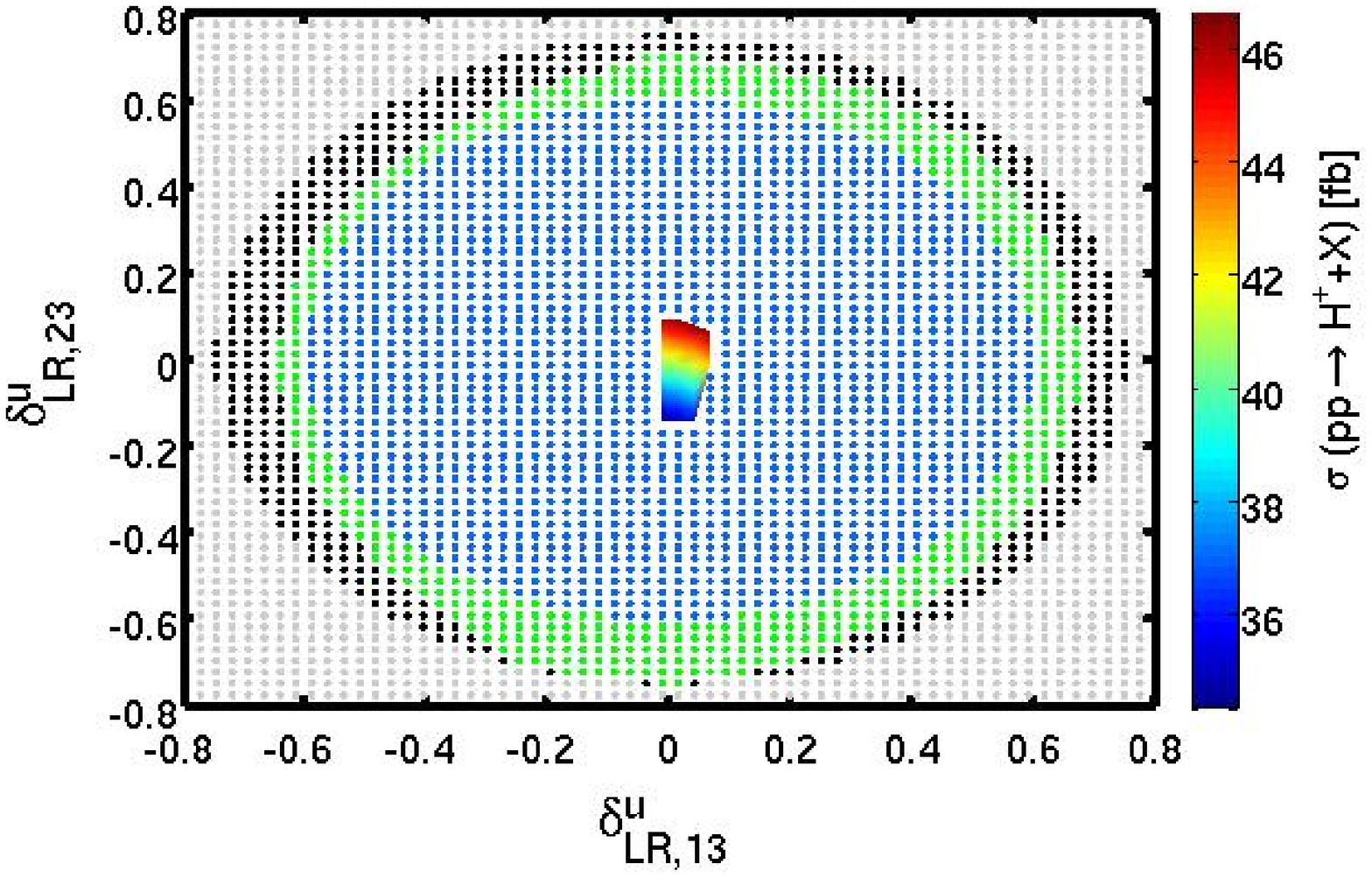} 
   \includegraphics[width=0.24\textwidth,height=0.22\textwidth,angle=0]{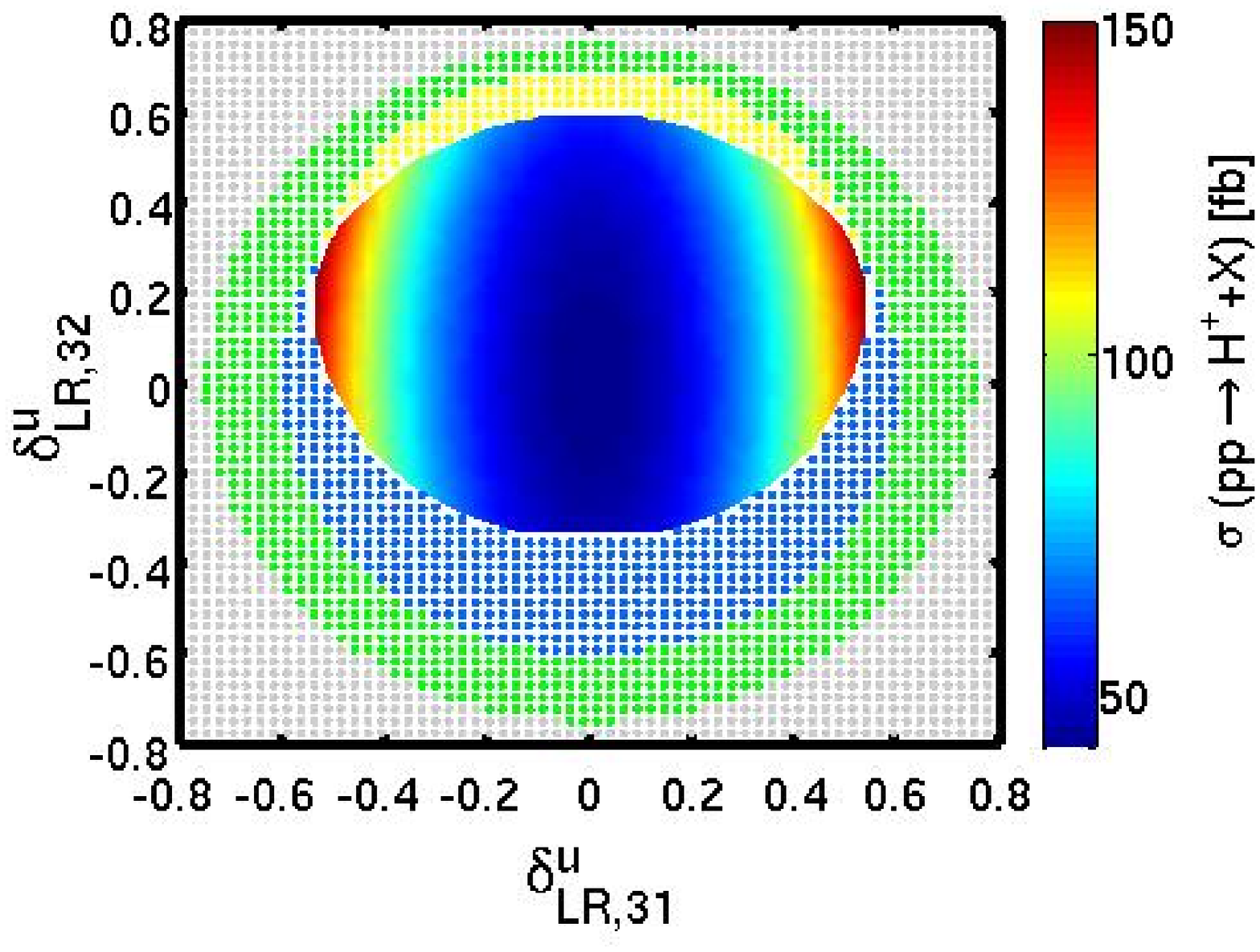} 
   \includegraphics[width=0.24\textwidth,height=0.22\textwidth,angle=0]{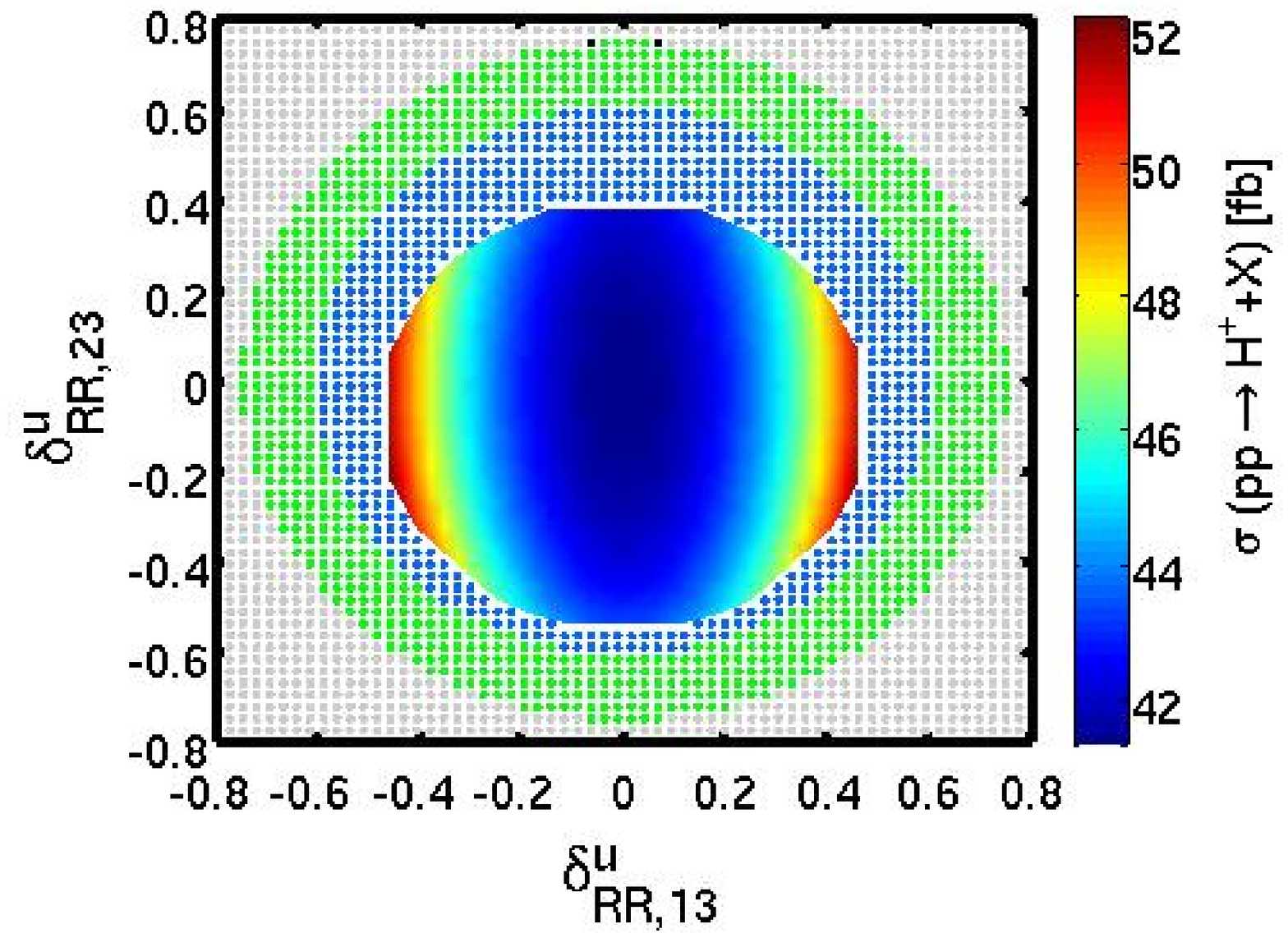}
 \end{center}
 \caption{Single-charged-Higgs production rates at the LHC. In the
 rainbow-colored area we include beyond-MFV parameters around the
 parameter point Table \ref{mytab}. Two
 $\delta_{AB,ij}^u$ are varied in each panel, all others are set to
 zero. The area outside the rainbow is ruled out experimentally}  
\label{fig:single_scan1}
\end{figure}

To test the effects of flavor structures on the single-Higgs
cross section we start with a generic MFV
SUSY parameter point which
does not violate any current bounds. We then allow for flavor
violation beyond MFV by $\delta_{AB}^q$. Because of current experimental and theoretical 
constraints discussed in Section~\ref{sec:constraints},
the up-squark parameters $\delta^u_{LR, 3i}$ and  $\delta^u_{RR, 3i}$
involving 1-3 and 2-3 mixing 
are the least constrained and therefore expected to 
cause the biggest effects. 

\begin{table}[tbp]
\begin{center}
\begin{tabular}{lll}
$\tan\beta = 7$ & $m_A = 170~\mathrm{GeV}$ & $\mu = -300~\mathrm{GeV}$   \\
$m_{\tilde q_{AA} \, ii} = 600~\mathrm{GeV}$ & $ M_2 = 700~\mathrm{GeV}$ & $m_{\tilde{g}} = 500~ \mathrm{GeV}$ \\ 
$A^{u,c} = 0$ & $A^{d,s,b} = 0 $ & $A^t = 1460~\mathrm{GeV}$ \\
\end{tabular}%
\end{center}
\caption{Generic parameter point}
\label{mytab}
\end{table}
Our starting parameter point is given in Table \ref{mytab}.
$m_A$ denotes the mass of the CP-odd Higgs leading 
to $m_{H^+}= 188~\mathrm{GeV}$. $M_2$ is the SUSY-breaking wino mass.
The diagonal soft-breaking entries in the squark mass matrices we choose
universal. All parameters are given at a scale of order $m_{H^+}$.
The large value of $A^t$ (corresponding to $\delta^u_{LR,33}$) increases the light Higgs mass
to $119.9~\mathrm{GeV}$ at two loops.
The production cross sections as a function of the three dominant
beyond-MFV mass insertions in the up-sector are shown in
Figure~\ref{fig:single_scan1}. Beyond-MFV effects can enhance the
single-Higgs rate to values above $100~\mathrm{fb}$. The size of the production cross
section is encoded in the rainbow scale in all panels of
Fig.~\ref{fig:single_scan1}, while the parameter choices outside this
area are ruled out. 
The different experimental
constraints impacting the parameter point shown in
Fig.~\ref{fig:single_scan1} include:
\begin{itemize} 
\item[--] Tevatron searches rule out the yellow points.
\item[--] squark searches and radiative and semileptonic decay limits
rule out the green points. 
\item[--] black points are forbidden by the squark--mass limits, $B$ mixing, and radiative and semileptonic decays. 
\item[--] blue points indicate a
violation of the radiative and semileptonic decay bounds only.
\item[--] orange points correspond to a violation of the $B$ mixing and radiative and semileptonic decay limits.
\item[--] grey points on the outside of the panels indicate a negative squark
mass square after diagonalizing the squark mass matrix. 
\end{itemize}

In Fig.~\ref{fig:single_scan1} we see that the limits on radiative
  and semileptonic decays followed by the 
Tevatron limit on light-flavor 
squark masses define two distinct boundaries of
forbidden parameter space.
After taking into account all limits, the off-diagonal
entry $\delta^u_{LR,31}$ has the strongest impact on the rate. 
It yields a maximal single-Higgs rate at
$|\delta^u_{LR,31}| \sim 0.6$.

\section{Charged-Higgs Production with a hard Jet \label{sec:hardjet}}

\begin{table}[tbp]
\begin{center}
\begin{tabular}{|cc|lll|}
\hline
$m_{H^+}$ & $\tan\beta$ & $\sigma_{\mathrm{2HDM}}$  & $\sigma_{\mathrm{MFV}}$ & $\sigma^{(m_f=0)}_{\mathrm{MFV}}$ \\[1mm] \hline
188~\textrm{GeV} & 3 & $2.5 \cdot 10^{-1}$  & $2.6 \cdot
10^{-1}$  & $6.7 \cdot 10^{-4}$ \\ 
188~\textrm{GeV} & 7 & $9.9 \cdot 10^{-1}$  & $1.1 \cdot 10^{0}$ & $1.5 \cdot 10^{-4}$ \\[2mm] 
400~\textrm{GeV} & 3 & $4.0 \cdot 10^{-2}$  &  $4.2 \cdot 10^{-2}$ & $4.2 \cdot 10^{-4}$ \\ 
400~\textrm{GeV} & 7 & $1.6 \cdot 10^{-1}$  & $1.7 \cdot 10^{-1}$  & $9.1 \cdot 10^{-5}$ \\ \hline
\end{tabular}%
\vspace{0.5mm}
\begin{tabular}{|cc|ll|}
\hline
$m_{H^+}$ & $\tan\beta$ & $\sigma_{\mathrm{SUSY}}$ & $\sigma_{\mathrm{SUSY}}^{(m_f=0)}$ \\[1mm] \hline
188~\textrm{GeV} & 3 & $14.3
\cdot 10^{0}$ & $13.9 \cdot 10^{0}$ \\ 
188~\textrm{GeV} & 7 & $4.6 \cdot 10^{0}$ & $3.0 \cdot 10^{0}$ \\[2mm] 
400~\textrm{GeV} & 3 & $2.4 \cdot 10^{0}$ & $2.3 \cdot 10^{0}$ \\
400~\textrm{GeV} & 7 & $7.9 \cdot 10^{-1}$ & $5.4 \cdot 10^{-1}$ \\ \hline
\end{tabular}%
\end{center}
\caption{Production rates (in fb) for the associated production of a charged Higgs with a hard jet: $p_{T,j}>100 \mathrm{GeV}$. The label 2HDM denotes a two-Higgs-doublet of type~II, while MFV and SUSY refer to the complete set of supersymmetric diagrams, assuming MFV and beyond. Beyond MFV we choose $\protect\delta^u_{LR,31} = 0.5$.}
\label{tab:hj}
\end{table}

The generic chiral suppression that characterizes single-Higgs production and
limits the cross section at tree level can be removed by simply adding an external
gluon to the operator basis.
Such operators can be of the
form $i\,\overline Q \gamma_\mu Q \, H_u
\leftrightarrow{D^\mu} H_u^C$, leading 
to higher-dimensional $q \bar q'Hg$ operators after electroweak
symmetry breaking \cite{operators}. 
To probe such operators at the LHC, we study charged-Higgs searches in
association with a hard jet. Simple diagrams for this process can be
derived from all single-Higgs production diagrams just radiating an
additional gluon. These are infrared divergent, which is no problem
once we require a hard jet with a typical $p_{T,j}>100~\textrm{GeV}$.

Similar to single-Higgs production we are interested in supersymmetric
loop corrections in and beyond MFV. We know from single-Higgs production that the flavor effects we
are interested in can be much larger than we expect next-to-leading
order QCD effects to be. Therefore, we ignore all gluonic
next-to-leading order corrections to charged-Higgs production with a
hard jet and limit our analysis to tree-level rates in the
two-Higgs-doublet model and additional supersymmetric one-loop
corrections with and without the MFV assumption. 

\subsection{MFV and beyond}

Assuming MFV, $F$-term and $A$-term
couplings of the Higgs to two squarks are proportional to the quark
masses, which means that supersymmetric one-loop amplitudes are
expected to be of the size of typical supersymmetric NLO
corrections. In the first column of Table~\ref{tab:hj} we list the 
hadronic tree level cross sections
for charged Higgs plus jet production 
for a non-supersymmetric two-Higgs-doublet type-II model. Numerical results for hadronic charged Higgs plus jet production in MFV 
we present in the second column in Table~\ref{tab:hj}. 

\begin{figure}[t]
 \begin{center}
   \includegraphics[width=0.24\textwidth,height=0.22\textwidth,angle=0]{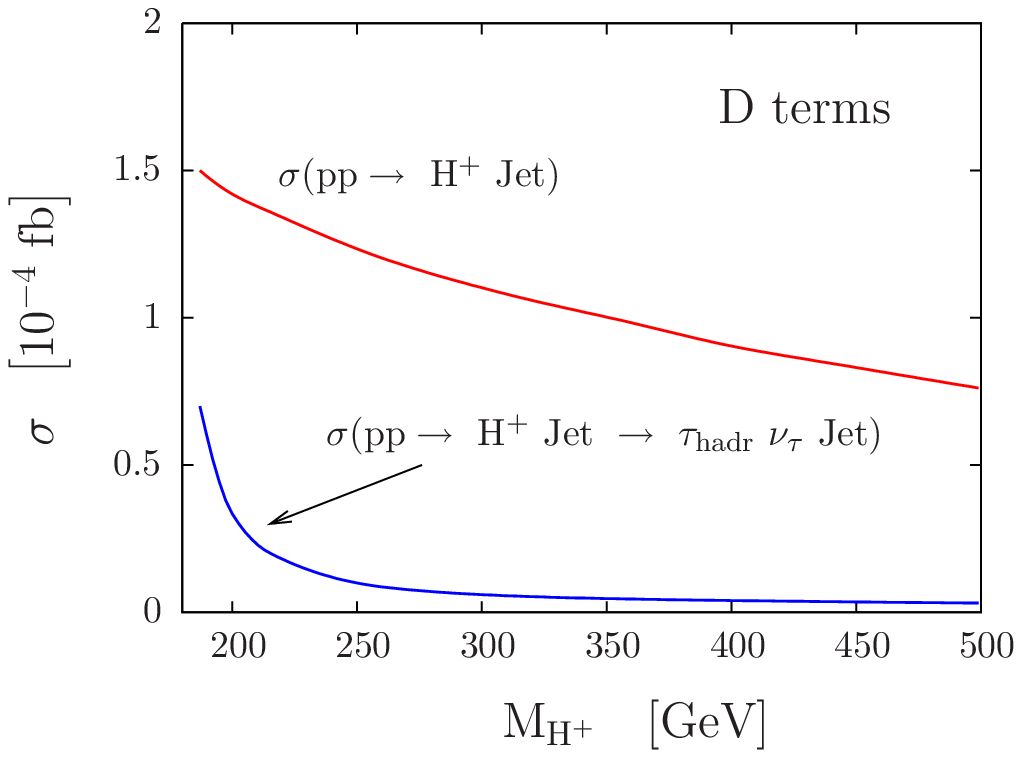} 
   \includegraphics[width=0.24\textwidth,height=0.22\textwidth,angle=0]{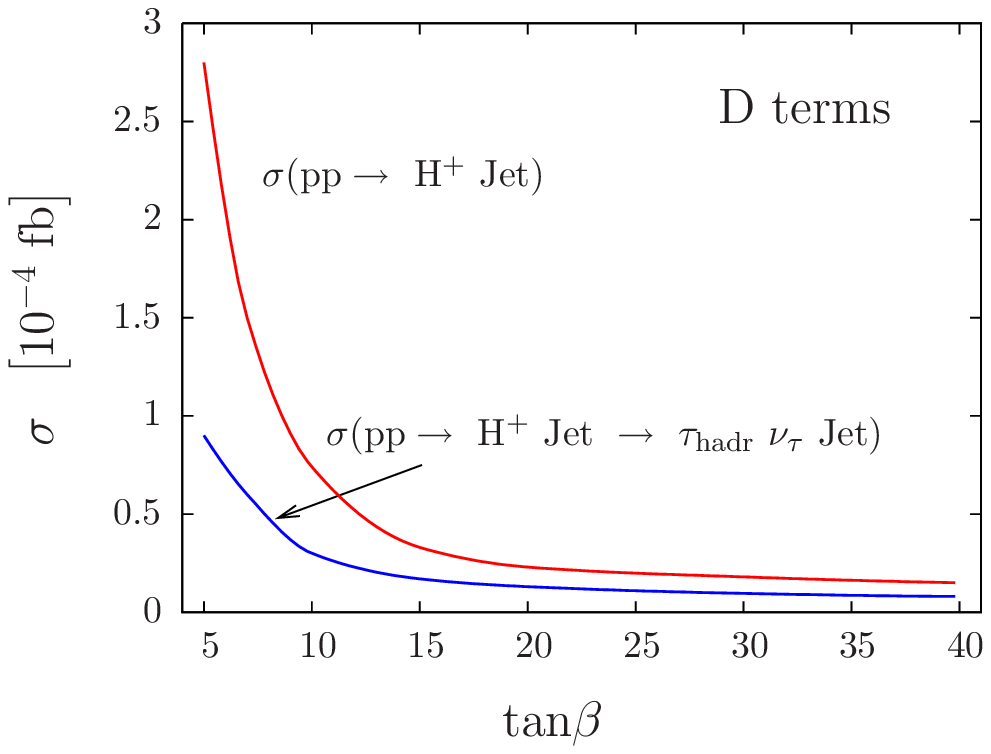} 
   \includegraphics[width=0.24\textwidth,height=0.22\textwidth,angle=0]{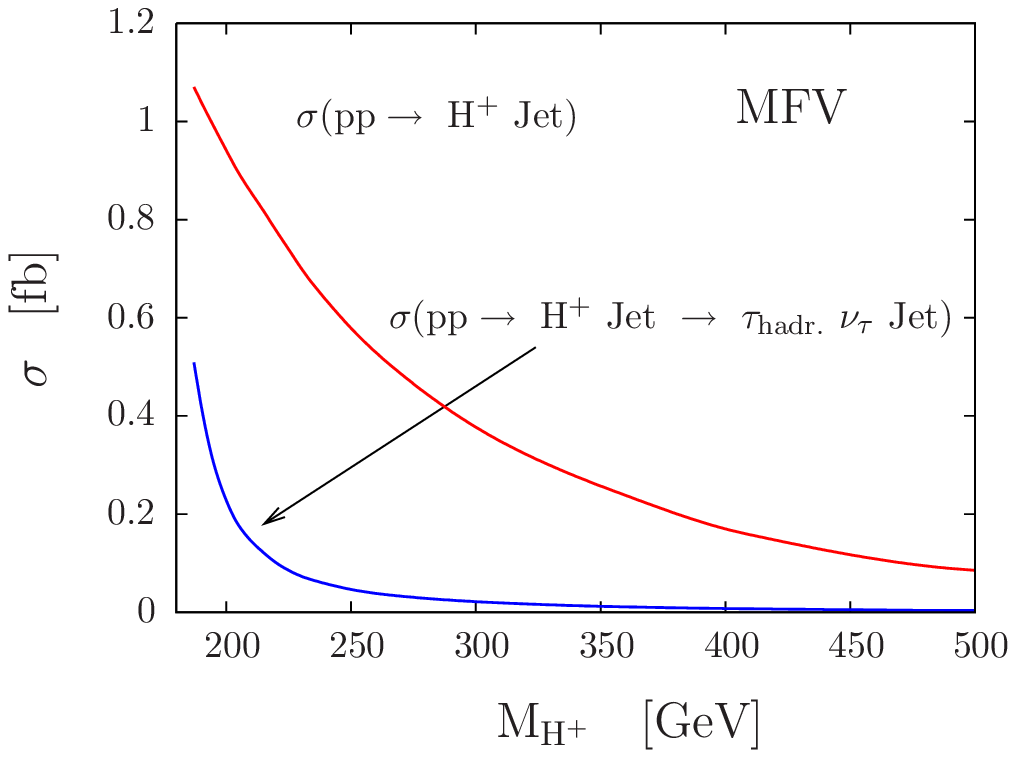} 
   \includegraphics[width=0.24\textwidth,height=0.22\textwidth,angle=0]{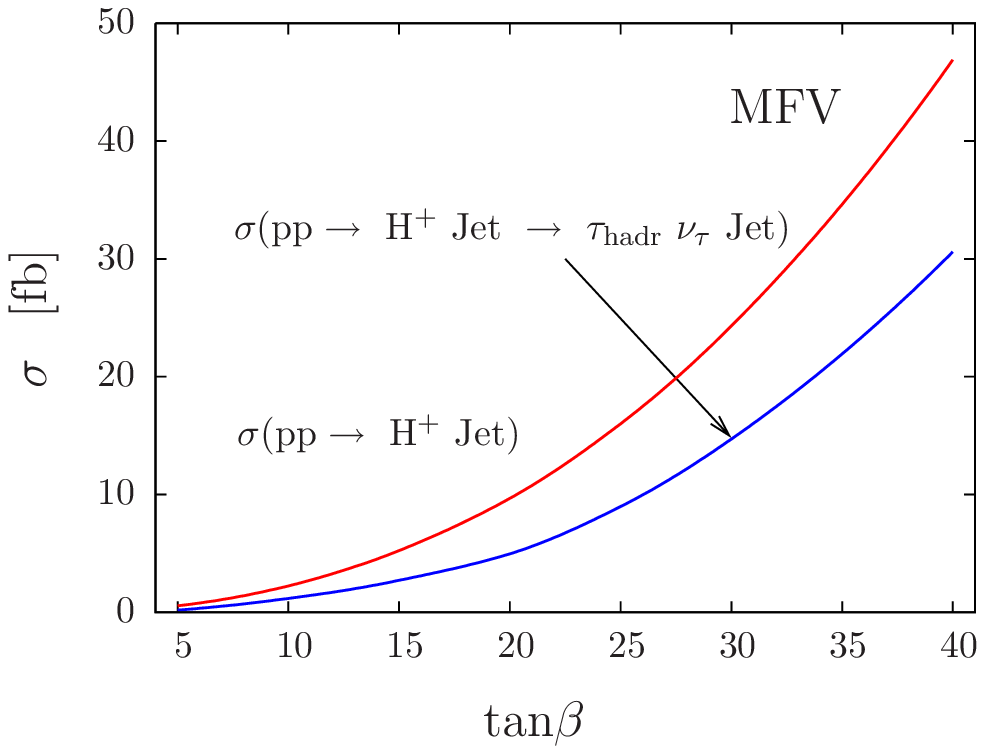}
 \end{center}
 \caption{Production rates for a
  charged Higgs with a hard jet including SUSY loops in MFV, assuming $m_f \rightarrow 0$ (upper panel) and $m_f \neq 0$ (lower panel).}
 \label{fig:asso_mfv}
\end{figure}

In the limit $m_f \rightarrow 0$ just the $D$-terms do contribute. Although chirally not suppressed and enhanced for small $\tan \beta$, the $D$-term
contribution is only a small fraction of the supersymmetric
amplitude (see Fig. \ref{fig:asso_mfv} and Tab. \ref{tab:hj}), due to its faster decoupling with heavy superpartner masses $\sigma \propto 1/M_{\rm SUSY}^8$.

In Fig.~\ref{fig:asso_mfv} we include Higgs decays as indicated and vary $\tan \beta$ and $m_{H^+}$. As long as the Higgs mass is small, $m_{H^+} \leq 200~\mathrm{GeV}$, the Higgs decay into a hadronic $\tau$ lepton is the most promising~\cite{dec_tau_ex}. The rates drop dramatically for heavier Higgs 
masses, even worse once we include the Higgs decay. Furthermore, with $m_f \neq 0$, the rates are not enhanced for small $\tan \beta$. The dominant background to this signature is clearly $W$+jet production $\sigma(pp \rightarrow W^+~\mathrm{Jet}) \approx 1~\mathrm{nb}$, with the $W$ decaying to a hadronic $\tau$.

Possible large supersymmetric corrections in this process can only occur 
beyond MFV --- just
like for single-Higgs production.

Although the operator basis does not get significantly extended when we include beyond-MFV effects, the effective couplings will get enhanced once we allow for sizeable $\delta^u_{AB,ij}$.
We respect the results form Section \ref{sec:singleH}, where we found $\delta^u_{RL, 13}$ to amplify the charged-Higgs cross section most. In the lower pattern of Tab.~\ref{tab:hj} we see, that, independent of the Yukawa couplings, beyond MFV can enhance the rate in the region of small $\tan \beta$ significantly, compared to the tree-level or MFV process. 
\section{Conclusion \label{sec:outlook}}
We find that if we allow for general squark mixing the cross sections for single-charged-Higgs production and charged-Higgs production in association with a hard jet can be enhanced by an order of magnitude, even after including all current experimental bounds.

The dominant source of genuine
supersymmetric flavor enhancement in the charged-Higgs production rate is the 
soft-breaking 
$A$ term for up-type squarks $A^u_{i3}$, which is invisible to kaon, charm and $B$-experi\-ments.
Hence, collider searches for enhanced charged-Higgs production rates can probe a unique sector of flavor.
A discovery would besides a breakdown of the Standard Model also signal
a non-standard solution to the flavor puzzle beyond the minimal-flavor-violation hypothesis.
Unfortunately, at present, we cannot firmly claim that these
flavor-induced charged-Higgs production rates at small $\tan\beta$
rates lead to observable signals over the large $W$-production
backgrounds.


\end{document}